# Pairwise, Magnitude, or Stars: What's the Best Way for Crowds to Rate?

**Alessandro Checco, Gianluca Demartini**
University of Sheffield, Sheffield, UK

## Abstract

We compare three popular techniques of rating content: the ubiquitous five star rating, the less used pairwise comparison, and the recently introduced (in crowdsourcing) magnitude estimation approach. Each system has specific advantages and disadvantages, in terms of required user effort, achievable user preference prediction accuracy and number of ratings required. We design an experiment where the three techniques are compared in an unbiased way. We collected 39'000 ratings on a popular crowdsourcing platform, allowing us to release a dataset that will be useful for many related studies on user rating techniques.

## Introduction

Users rating content on the Web is a key activity for a variety of applications: from recommender systems to information retrieval system evaluation. The most common way for users to rate content is star rating. In 2006, Netflix released a dataset containing 100 million movie ratings using the star system, offering a $1M prize to improve their recommender system (Bennett and Lanning 2007).

Alternatives to the star rating approach exist. For example, magnitude estimation, originally developed for psychophysical measurement (Stevens 1966), has been recently proposed for crowdsourced ratings collection applied to information retrieval evaluation (Turpin et al. 2015). With this method users are allowed to use any numerical value to rate content so that they are always free to put an higher/lower score as compared to the content they have seen so far.

Pairwise comparison has a long history, but it has the problem of requiring a high number of comparisons to achieve good user preference prediction accuracy (Wauthier, Jordan, and Jojic 2013).

## Experimental Setup

The experiment design we use to compare these three rating approaches is structured as follows. We selected 10 most popular images of paintings, obtained from artcyclopedia.com top 10 poster sales. We then asked crowd workers to rate them using three different methods:

**Magnitude** Using any positive integer (zero excluded).

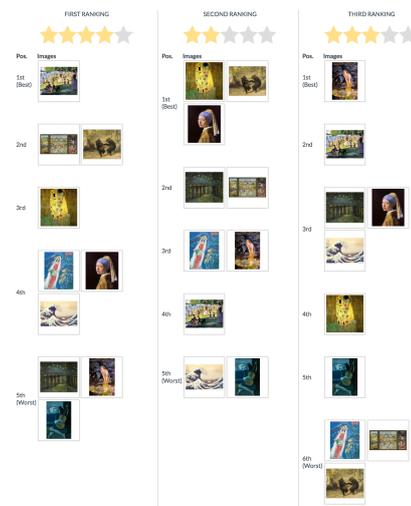

Figure 1: Graphical interface to let the worker express their preference on the ranking induced by their own ratings.

**Stars** Choosing between 1 to 5 stars.

**Pairwise** Pairwise comparisons between two images, with no ties allowed (note that this requires 45 comparisons for our 10 images!).

We ask each worker to rate the images using all 3 rating systems. Since the order with which we ask to use a different rating system affects the outcome, we run 6 different experiments (one for each combination of these three types) with 100 participants in each of them. We can thus analyse the bias given by the rating system order, and the results without order bias by using the aggregated data. We obtain a dataset with a total of 39'000 ratings (45+10+10)[1].

At the end of the rating activity in the task we dynamically build the three painting rankings induced by the choices of the participant (pairwise ranking is obtained by a one point tournament), and ask them which of the three rankings better reflects their preference[2] (an example screenshot is shown in Figure 1). This will reveal which kind of rating system is preferable. We also collect the time spent in each rating activity.

---

[1] The dataset is available for download at https://github.com/AlessandroChecco/PairwiseMagnitudeStars

[2] The ranking comparison is blind: There is no indication on

|            | Mean | Median | Preferences |
|------------|------|--------|-------------|
| Magnitude  | 3.74 | 4.0    | 98          |
| Stars      | 3.89 | 4.0    | 107         |
| Pairwise   | 4.30 | 5.0    | 243         |

Table 1: Mean and median rating (out of 5) of the ranking induced by the three techniques. We also report the number of times a method was preferred by workers over the others (excluding ties).

## Results

In Table 1 we see that participants clearly prefer the ranking obtained from their pairwise comparisons. A binomial test shows that the difference in preferences is statistical significant ($p < 1E-19$).

In Figure 2 the preferred technique count is shown, now grouped by the different test orders (where the techniques are abbreviated by their initials). We notice a memory bias effect: The last technique used is more likely to get the most accurate description of the real user preference. Despite this, the pairwise comparison technique obtained the maximum number of preferences in all cases.

In Figure 3, we show the average time that participants needed to complete the rating activities, grouped by the order in which the tests have been run. We also show the theoretical value that could be achieved with a dynamic test system for pairwise comparison (of order $N \log N$ as proved in (Wauthier, Jordan, and Jojic 2013)). While, as expected, the pairwise comparison technique clearly requires more time than the other techniques (with our exhaustive approach it required $N(N-1)/2$ comparisons), we can see that it would be comparable in terms of time with the other techniques using less comparisons. We can argue that using additional information (like common patterns and clustering of user preferences) even better results could be achieved. This question is important and left for future research, where this dataset will be useful to obtain unbiased analyses.

The magnitude technique is clearly the worst in term of time taken per question while star and pairwise rating are much more efficient, indicating that the effort required for magnitude estimation is considerably higher than the other techniques. This is corroborated by free text user feedback, where we observed that many users find difficult to decide and adjust the scale of the magnitude technique, requiring to remember the whole previous rating set.

## Conclusions

Star rating is confirmed to be the most familiar way for users to rate content, whereas magnitude estimation has revealed to be unintuitive with no added benefit. Overall, pairwise comparison, while requiring a slightly higher number how each ranking has been obtained, and their order is randomized.

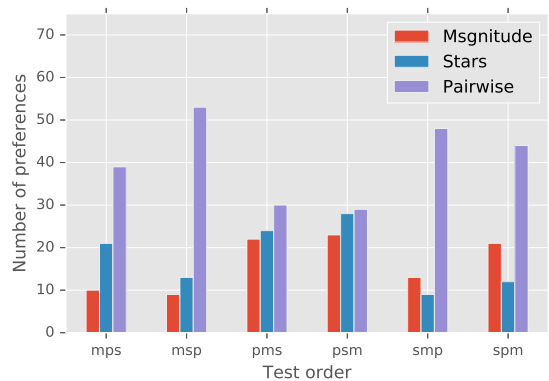

Figure 2: Number of expression of preference of the ranking induced by the three different techniques, grouped by the order in which the tests have been run.

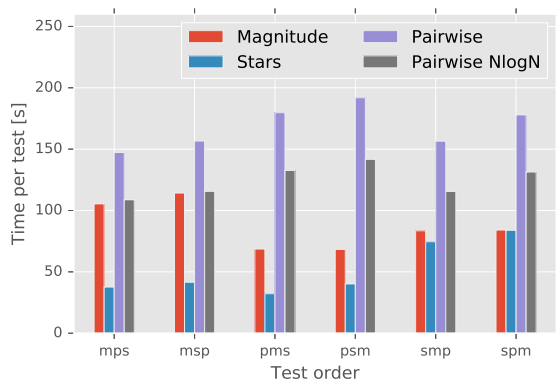

Figure 3: Average time per test, grouped by the order in which the tests have been run. Also the theoretical value that could be achieved with a dynamic test system for pairwise comparison is shown.

of low-effort user ratings, best reflects intrinsic user preferences and seems to be a promising alternative to star rating.

## References

[Bennett and Lanning 2007] Bennett, J., and Lanning, S. 2007. The netflix prize. In *Proceedings of KDD cup and workshop*, volume 2007, 35.

[Stevens 1966] Stevens, S. S. 1966. A metric for the social consensus. *Science* 151(3710):530–541.

[Turpin et al. 2015] Turpin, A.; Scholer, F.; Mizzaro, S.; and Maddalena, E. 2015. The benefits of magnitude estimation relevance assessments for information retrieval evaluation. In *Proceedings of the 38th International ACM SIGIR Conference on Research and Development in Information Retrieval*, SIGIR '15, 565–574. New York, NY, USA: ACM.

[Wauthier, Jordan, and Jojic 2013] Wauthier, F. L.; Jordan, M. I.; and Jojic, N. 2013. Efficient ranking from pairwise comparisons. *ICML (3)* 28:109–117.